\documentclass[journal]{IEEEtran}

\usepackage{multirow}
\usepackage{amsfonts,amsmath,amssymb,mathtools}
\usepackage{graphicx}
\usepackage{epstopdf}
\usepackage{xurl}
\usepackage{cite}
\usepackage[caption=false,font=footnotesize,subrefformat=parens,labelformat=parens]{subfig}
\usepackage{comment}
\usepackage[ruled,vlined]{algorithm2e}
\usepackage{algpseudocode}
\usepackage{tikz}
\usetikzlibrary{arrows,intersections,decorations,plotmarks,math,patterns}
\usepackage{pgfplots}
\pgfplotsset{compat=1.17}
\usepackage{color}
\usepackage[colorinlistoftodos]{todonotes}

\newcommand\scalemath[2]{\scalebox{#1}{\mbox{\ensuremath{\displaystyle #2}}}}

\definecolor{COL11}{RGB}{215,25,28}
\definecolor{COL12}{RGB}{253,174,97}
\definecolor{COL13}{RGB}{146,197,222}
\usepackage{hyperref}
\usepackage{booktabs} 

\begin{document}

\title{Spatio-Temporal Consistent Soft Sensor Modeling and Monitoring of Thermal Power Plants based on Physical Knowledge}

\author{
Qianchao Wang$^{1}$, Peng Sha $^{2}$, Leena Heistrene$^{3}$, Yuxuan Ding$^{1\dag}$, Yaping Du $^{1}$,  

\thanks{Yuxuan Ding is the corresponding author}
\thanks{$^{1}$ Qianchao Wang, Yuxuan Ding, and Yaping Du is with the Department of Building Environment and Energy Engineering, Hong Kong Polytechnic University, Hung Hom, Hong Kong. (qianchao.wang@polyu.edu.hk; yx.ding@connect.polyu.hk; Ya-ping.du@polyu.edu.hk )}
\thanks{$^{2}$ Peng Sha is with Key Laboratory of Energy Thermal Conversion and Control of Ministry of Education, School of Energy and Environment, Southeast University, Nanjing, China. (sha\_peng@seu.edu.cn) }
\thanks{$^{3}$ Leena Heistrene is with Department of Electrical Engineering, School of Energy Technology, Pandit Deendayal Energy University, Gandhinagar, Gujarat, India. (leena.santosh@sot.pdpu.ac.in) }

}

\maketitle
\begin{abstract}
Data-driven soft sensors have been widely applied in complex industrial processes. However, the interpretable spatio-temporal features extraction by soft sensors remains a challenge. In this light, this work introduces a novel method termed spatio-temporal consistent and interpretable model (STCIM).
First, temporal and spatial features are captured and aligned by a far topological spatio-temporal consistency extraction block. Then, the features are mapped into an interpretable latent space for further prediction by explicitly giving physical meanings to latent variables. The efficacy of the proposed STCIM is demonstrated through the modeling of two generated datasets and a real-life dataset of coal-fired power plants. The corresponding experiments show: 1) The generalization of STCIM outperforms other methods, especially in different operation situations. 2) The far topological spatio-temporal consistency is vital for feature alignment. 3) The hyper-parameters of physics-informed interpretable latent space loss decide the performance of STCIM.

\end{abstract}

\section{Introduction}

\IEEEPARstart{T}{o} reduce carbon emissions and mitigate the greenhouse effect, the modeling and monitoring of thermal power plants become the cornerstones of stable operation and security in the energy system. However, data offset, noise disturbance, low reliability of analyzers, and delays in long time-series data from most processes lead to a mismatch between models and the data, resulting in potential safety hazards and poor product quality. As a solution, soft sensors are used for process monitoring, quality prediction, and many other important applications in thermal power plants~\cite{9235582}.

There are two main types of approaches to establish soft sensing models, namely, mechanism-based methods~\cite{Huang2013DynamicMA} and data-driven methods~\cite{10329481,9794453}. Mechanism-based methods, also called first-principle models, use physicochemical governing equations and dynamic differential equations to describe industrial processes~\cite{mukherjee2025development,van2025component}. It can model or monitor thermal power plants in an interpretable and reliable way. When we have sufficient knowledge about thermal power plants and the process, mechanism-based methods can work effectively, improving practitioners’ trust in the models. However, the complexity of thermal power plants often makes these preconditions difficult to meet, limiting the applicability of the first principles. In contrast, data-driven methods, particularly deep learning (DL), have achieved significant success in chemical, biochemical, and metallurgical processes, since they do not require extensive prior knowledge~\cite{8996997,9329169,10967531}.

In recent research, deep learning methods have been increasingly applied to large-scale industrial processes, with a particular emphasis on spatio-temporal feature extraction to capture both spatial dependencies and temporal dynamics among variables~\cite{10530070}. These approaches build on foundational techniques such as multiscale information fusion~\cite{10648819} and dynamic feature extraction~\cite{lui2022supervised}, but prioritize spatio-temporal integration to address complex interactions. Some efforts focused on local spatio-temporal features~\cite{10845809}, incorporating mechanisms like BiGRU with attention for time-related patterns~\cite{10926880} and dynamic convolutional neural networks with multiscale attention for hierarchical local nonlinear dynamics~\cite{10465636}. However, the far topological structures in industrial variables often limit local methods, prompting advancements in broader spatio-temporal modeling~\cite{11021627}. For instance, variable correlation analysis-based convolutional neural networks have been proposed to extract far topological spatio-temporal features in processes like hydrocracking and debutanizer columns~\cite{10458995}. Similarly, spatio-temporal attention combined with long short-term memory networks identifies long-term variable correlations in hydrocracking~\cite{9062588}, while graph neural networks process relational structures for enhanced spatio-temporal feature extraction~\cite{yang2025akgnn}. However, many learned broader spatio-temporal features assume alignment across space and time, overlooking delays and couplings among variables, which can lead to reduced generalization in real-world applications. This problem is particularly evident in thermal power plants and needs to be solved.

Another problem is that black-box models lack physical interpretability, critical for safety in thermal power plants. Sparse spatio-temporal features in long time-series data are not explainable by first principles, limiting their utility. While graph neural networks incorporate prior knowledge to identify physical correlations~\cite{10129979,10530070}, they lack explicit physical grounding. Physics-informed neural networks (PINNs) address this by embedding physical laws into loss functions via differential equations and initial conditions, leveraging automatic differentiation for transparency~\cite{RAISSI2019686,pittphilsci23067,9743327}. Successful applications include lake temperature prediction~\cite{daw2021physicsguided}, parameter estimation~\cite{9647978}, and flexibility analysis~\cite{9771395}. Despite advantages in efficiency, noise immunity, and accuracy, PINNs implicitly encode physical properties, lacking explicit expressions of these relationships.

Considering this gap, the primary objective of this study is to introduce an explicitly interpretable spatio-temporally consistent soft sensor model in thermal power plant modeling and monitoring.
We propose a spatio-temporal consistent and interpretable model (STCIM) with a spatio-temporal consistent feature extraction block and 
an interpretable latent space. This method unfolds in two distinct phases: in the first stage, a root layer is designed to extract the local features at different times, and a trainable position encoding layer is embedded into the root layer to ensure the local spatio-temporal consistency. Subsequently, the local consistent features are utilized as the input to a far topological alignment layer, computing the correlations among the features by attention mechanism. During the process, the position encoding layer can adaptively update itself to ensure the consistency of far topological spatio-temporal features. 
In the second stage, the spatio-temporal consistent features are projected into the latent space. The latent space is guided by the first principle, explicitly assigning physical meanings to maintain the interpretability of the model. In training process, the latent space is integrated into the loss function to predict the state variables and the output.
Notably, this methodology holds potential for adaptation to other domains within industrial modeling or diagnostic applications. The efficacy of these concepts is validated through extensive experimentation using simulated datasets and real-world data from a 330MW thermal power plant. In summary, the key contributions of this paper encompass:
\begin{enumerate}
	\item A spatio-temporal consistent and interpretable model (STCIM) is proposed for thermal power plant modeling and monitoring. It efficiently extracts and aligns the spatio-temporal features and gives the latent variables physical meanings to improve the model's interpretability.

	\item The spatio-temporal consistent feature extraction block is introduced to extract and align local and global spatio-temporal features by using trainable position encoding and attention modules.
    
    \item The physics-informed interpretable latent space loss function is introduced to map the latent space into the state variables by giving explicit physical meanings to latent space and integrating them in a discrete state-space loss function. The discrete state-space loss function is a part of the physics-informed interpretable latent space loss function.

    \item  The efficacy of STCIM is underscored through its application experiments and ablation experiments in two simulated datasets and two real thermal power plant datasets, thus substantiating its effectiveness in practical settings. 
\end{enumerate}

The paper is organized as follows: Section~\ref{sec:background} provides background about the thermal power plant, basic blocks, and physics-informed loss functions. Section~\ref{sec:Method} describes the detailed information of the proposed STCIM, including the model structure, the spatio-temporal consistent feature extraction block, and physics-informed interpretable latent space loss function. Section~\ref{sec:Simul} shows the experimental results, and Section~\ref{sec:conclusion} concludes the paper. 

\section{Brief Introduction of Thermal Power Plants} \label{sec:background}

\begin{figure}[!t] 
\centering
\includegraphics[width=0.95\linewidth]{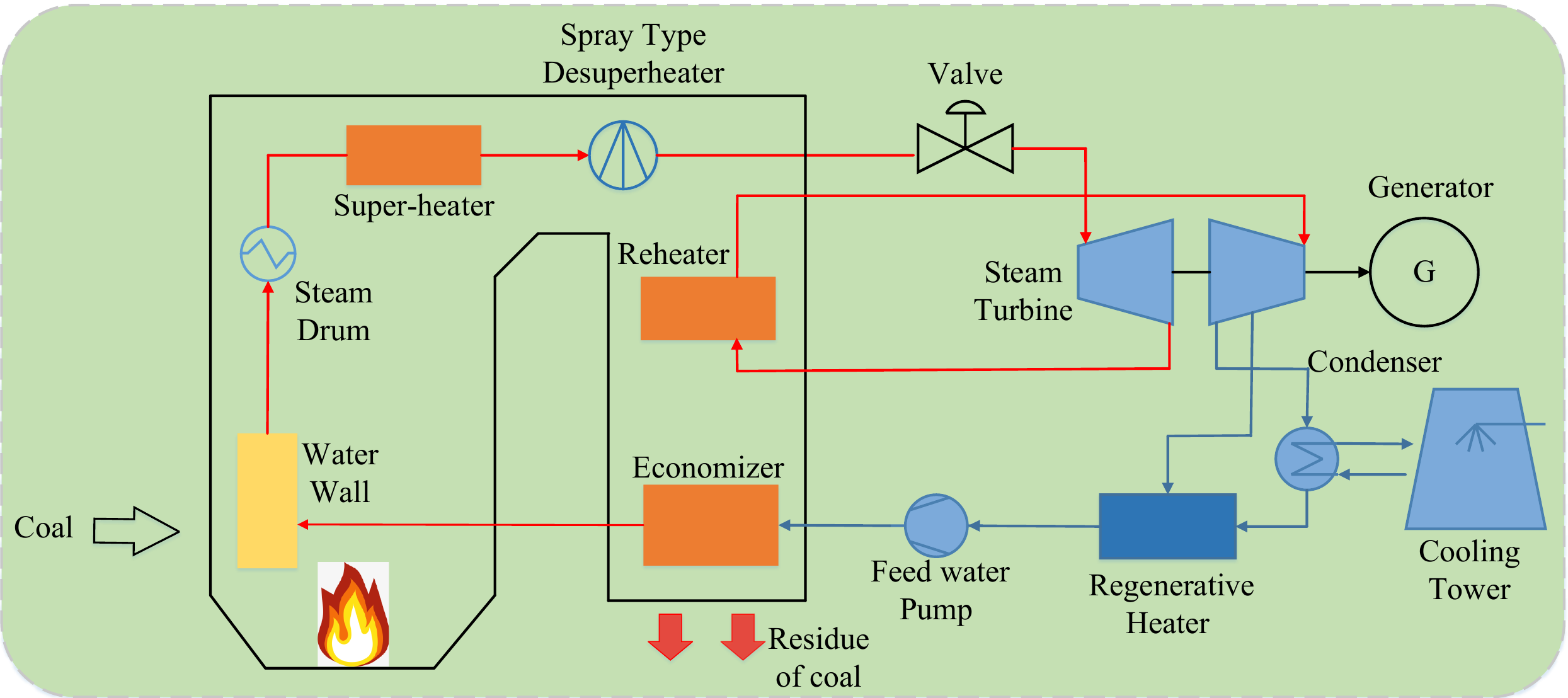}
\caption{330MW thermal power plant.}
\label{fig:330MW coal power plant}
\vspace{-0.3cm}
\end{figure}

Thermal power plant is a multifaceted and complex industrial process, generating electricity by converting heat energy from fossil fuels into electrical energy~\cite{fan2025dynamic}. It contains the chemical reactions, energy conversion, mass transfer which is shown in Fig.~\ref{fig:330MW coal power plant}. The process begins with fuel combustion in a boiler, producing high-pressure steam that drives a turbine connected to a generator, which produces electricity. After passing through the turbine, the steam is cooled in a condenser and recycled, while the exhaust gases are treated and released through a chimney.

The mathematical description of the utilized 330MW thermal power plant~\cite{WANG2024110800,wang2020improving} can be given by 
\begin{equation}\label{eq:coal}
{\begin{split}
&\acute{r_{B}} =e^{-18s}u_{B} \\
&120\frac{dr_{B}}{dt}=-r_{B}+\acute{r_{B}} \\
&3266\frac{dp_{d}}{dt}=-0.2501p_{t}u_{T}+6.77r_{B} \\
&12\frac{dN_{E}}{dt}=-N_{E}+0.2501p_{t}u_{T} \\
&p_{t}=p_{d}-0.0004555(6.77r_{B})^{1.3} \\
\end{split}}
\end{equation}
where $u_{B}$ is the fuel signal ($kg/s$), $u_{T}$ is the opening of turbine ($\%$), $r_{B}$ is the actual fuel ($kg/s$), $p_{d}$ is the pressure of steam (MPa) and $N_{E}$ is the output power (MW). $\dot{z_{1}} = \frac{dr_{B}}{dt}$, $\dot{z_{2}} = \frac{dp_{d}}{dt}$ and $\dot{y} = \frac{dN_{E}}{dt}$ ($z_{1}, z_{2}$ are the state variables). It is derived from the first principles of the thermal power plant.

\section{Spatio-Temporal Consistent and Interpretable Modeling} \label{sec:Method}

Modern multi-coupled industrial processes involve interconnected production units with complex material and energy transfers, posing challenges in modeling and monitoring due to time delays, variable coupling, and interpretability. We developed a trainable position coding to align spatio-temporal consistency during local and far topological feature extraction. Additionally, we designed a physics-informed interpretable latent space loss function by assigning physical meanings to latent variables and incorporating them into the loss function. This section details the STCIM, focusing on the spatio-temporal consistent feature extraction block and the physics-informed interpretable latent space loss function.

\begin{figure*}[!t] 
\centering
\includegraphics[width=0.95\linewidth]{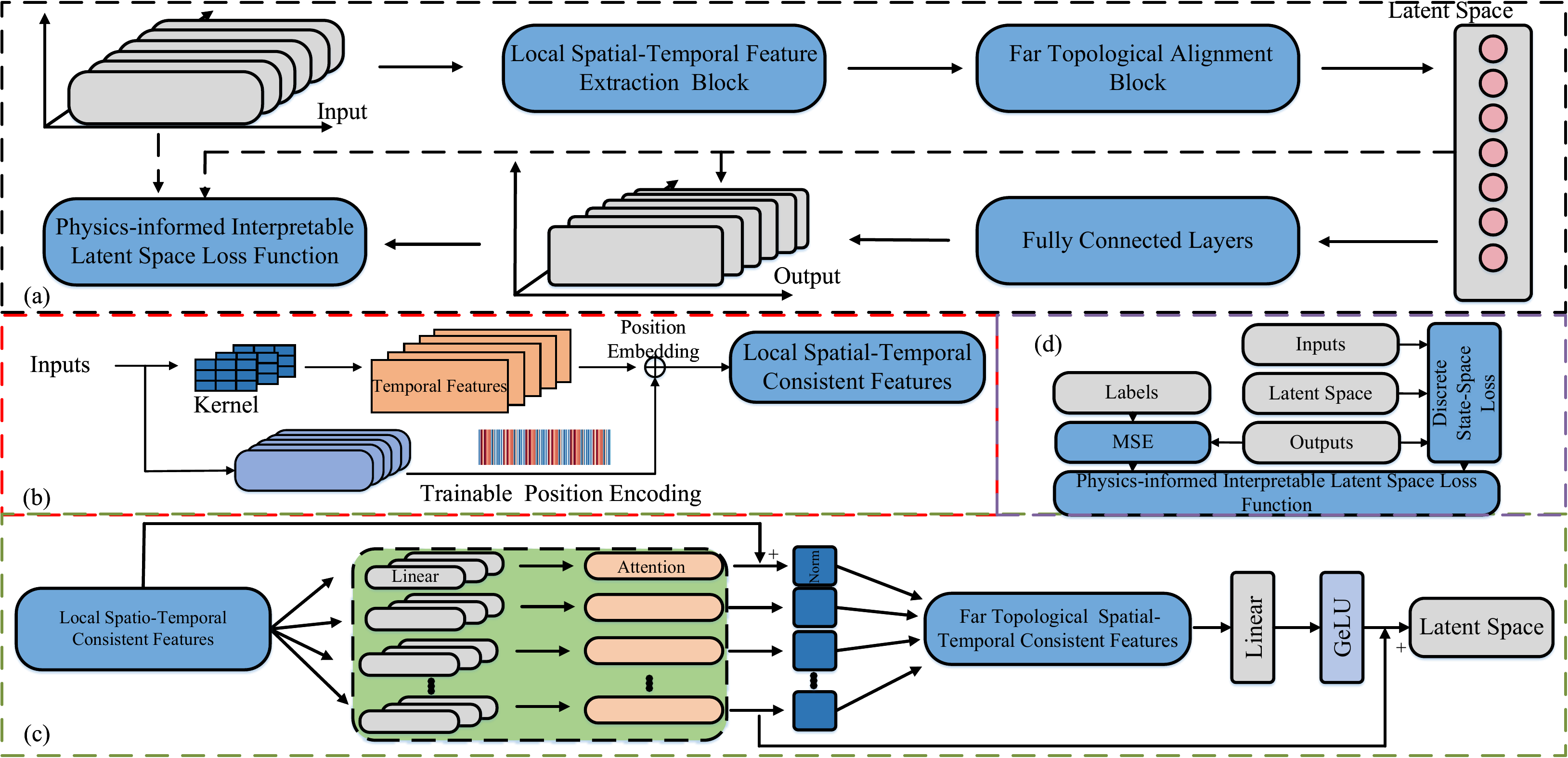}
\caption{Overview of the STCIM framework. (a) The flow chart of STCIM. (b) The local spatio-temporal feature extraction block. (c) The far topological alignment block. (d) The physics-informed interpretable latent space loss function}
\label{fig:Overview of the proposed STCIM}
\vspace{-0.3cm}
\end{figure*}

\subsection{Model Structure}

Fig.~\ref{fig:Overview of the proposed STCIM} (a) shows the flow chart of the proposed STCIM. It can be recognized as an encoder-decoder structure, in which the spatio-temporal consistent feature extraction block (encoder) is divided into two sub-blocks: the local spatio-temporal consistent feature extraction block and the far topological alignment block, and the decoder is a fully connected network. 

In the local spatio-temporal consistent feature extraction block, the temporal features are extracted by the kernels with different size. The trainable position layer is added in parallel to temporal feature extraction as the position embedding to ensure the consistency of local spatio-temporal features. Then the local features are utilized to calculate the far topological correlations based on the attention mechanism, capturing the coupling among features in the far topological alignment block. During the process, the position encoding can adaptively update itself based on the new captured coupling features to ensure the consistency of far topological spatio-temporal features. 

The latent space is another important node, which is utilized to explicitly add physical meanings to models. The spatio-temporal consistent features are projected into the latent space and constrained by the first principles by adding a discrete state-space loss in the physic-informed  interpretable latent space loss function. The size of the latent space is determined by the form of discrete state-space equations, e.g., if there are two state variables in a discrete model, then the number of latent variables for STCIM is four because it contains the state variables at time $k$ and $k-1$. The decoder is composed of MLPs, which is enough for discrete function fitting.

In general, the idea of the model can be summarized as compressing the spatio-temporal consistent features with high dimensions into an interpretable latent space, which normally has fewer dimensions, and then the interpretable latent variables are utilized to fit the potential mathematical equations, for example, the state-space equations. 



\subsection{Spatio-Temporal Consistent Feature Extraction Block} 


The spatio-temporal consistent feature extraction block mainly focuses on the consistency of space and time in features by designing a trainable position encoding layer. It contains the local spatio-temporal consistent feature extraction block and the far topological alignment block. In the two blocks, during the training process, the position encoding layer can adaptively update itself to ensure the spatio-temporal consistency in any features.

\subsubsection{Local Spatio-Temporal Feature Extraction Block}

Fig.~\ref{fig:Overview of the proposed STCIM} (b) describes the local spatio-temporal feature extraction block. The input data, respectively, pass through a temporal feature extraction layer and a trainable position encoding layer to capture the local features from different times and locations. The massive local spatio-temporal features ensure the abundance of far topological variables.

Assume that the input variables at time $t$ are $X(t)\in R^{H\times W\times C}$ where $H, W$, and $C$ are separately the backtracking from time t, the number of features, and the number of input channels ($C=1$). $X(t)$ can be expressed as
\begin{equation} \label{eq:inputs}
\scalemath{0.7}{X(t)= \begin{bmatrix}
  x_1(t-(H-1)\times T)& x_2(t-(H-1)\times T) & \cdots  & x_n(t-(H-1)\times T)\\
  \vdots & \vdots & \vdots & \vdots \\
  x_1(t-2\times T)& x_2(t-2\times T ) & \cdots & x_n(t-2\times T ) \\
  x_1(t-1\times T)& x_2(t-1\times T ) & \cdots & x_n(t-1\times T ) \\
  x_1(t)& x_2(t)  & \cdots   & x_n(t)
\end{bmatrix}}
\end{equation}
where $T$ is the sampling time. When different convolution kernels are applied in $X(t)$, the local features of adjacent attributes and times $S_{i,j}$ are extracted, reflecting the dynamics of the industrial process, $S_{i,j}=\sum_{m=1}^{m}\sum_{n=1}^{n}K_{m,n}X(t)_{i+m,j+n}+b$. 

However, this kind of one-shot input is disorder, ignoring the potential position information in attributes, which is opposite to the clear production procedures of the industrial process. Therefore, we designed a trainable position encoding layer, providing the corresponding spatial features to ensure the consistency of spatio-temporal features. Inspired by~\cite{dosovitskiy2021image}, the trainable position encoding layer is given by
\begin{equation} \label{eq:position encoding}
Position_{j}= \frac{Variable_{j}}{\sqrt{N} } 
\end{equation}
where $Variable_{j}$ is the trainable position parameter of $j_{th}$ attributes. $N$ is the number of features that is a scaling factor. $Position_{j}$ is the position encoding of the $j_{th}$ attributes. Due to the unknown time delay and coupling among variables in industrial processes, fixed position encoding is not appropriate, for example, 'sin' and 'cos' encoding. The trainable position parameter $Variable_{j}$ in this layer can adaptively ensure spatio-temporal consistency of captured local features.The final local spatio-temporal consistent features can be summarized as
\begin{equation} \label{eq:local}
\hat{S_{j}}= S_{j}+Position_{j}
\end{equation}
where $\hat{S_{j}}$ is the final feature of $j_{th}$ attributes.

\subsubsection{Far Topological Alignment Block}

During the industrial process, variables are related not only to adjacent nodes but also to variables in far topological structures. Aligning the far topological correlation among local features and making sure the spatio-temporal consistency are vital for the modeling and monitoring of thermal power plant. In this vein, we propose a far topological alignment block, which is shown in Fig.~\ref{fig:Overview of the proposed STCIM} (c).

The local spatio-temporal consistent features are first sparsely mapped by the multi-head linear  transformations to highlight different aspects of the features. Then several self-attention modules are utilized to capture the far topological structure in features by calculating the attention weights of all features. The calculated weights can be considered as the dependencies among far topological features and captured by models.
Subsequently, the captured topological and spatio-temporal features pass through a fully connected module with GeLU to provide nonlinear properties. In the forward process, there are two shortcut connections to ensure the flow of information. During the far topological variable correlation extraction process, the adaptive position encoding at the local spatio-temporal feature extraction layer is realigned with new features, which will again ensure the consistency of the far topological spatio-temporal feature and improve the modeling and monitoring accuracy.

Assume that the features captured by the local spatio-temporal feature extraction block are $\hat{S}\in R^{H\times W\times K}$.
Based on the attention weights, the dependencies among far topological features can be calculated by
\begin{align} \label{eq:position encoding}
& Weight(\hat{S},\hat{S})=SoftMax(\frac{\hat{S}\hat{S}^T}{\sqrt{d_k}})  \\
& S_{new}= Weight(\hat{S},\hat{S})\hat{S}
\end{align}
where $Q=K=V=\hat{S}$. The calculated weights determine the intrinsic correlation among the input features, helping the networks capture the potential far topological structure among all the features. By multiplying $Weight(\hat{S},\hat{S})$ with $\hat{S}$, the importance determined by the long-term topological correlation can be used in feature extraction.

\subsection{Physics-informed Interpretable Latent Space Loss Function} 
To map these sparse spatio-temporal consistent features into an interpretable latent space and limit the space in first principles, this subsection introduces the physics-informed interpretable latent space loss function by giving physical meanings to the latent space $z$ of STCIM and adding them to the discrete state-space loss, which is shown in Fig.~\ref{fig:Overview of the proposed STCIM} (d). We can directly relate the inputs, latent variables, and outputs by the discrete state-space loss function and boundary conditions. Here, we need to highlight that the latent space $z$ is an intermediate process parameter instead of the input of STCIM. It can be unobserved or hard to measure.

Assuming the discrete-time state-space equation and output equation of the industrial process are given by
\begin{align} \label{eq:state space  equations}
&u(t)=\Phi(u(t-1),\theta_{1})+G(x(t),\theta_{2})\\
&y(t) =H(u(t),\theta_{3})+J(x(t),\theta_{4})
\end{align}
where $x$ is the input, $u$ is the unobserved state variables, and $y$ is the output. $\Phi(\cdot ),G(\cdot ),H(\cdot ),J(\cdot )$ are separately the nonlinear operators. $\theta$ is the parameter. All variables are vectors. 

Now, we can map the discrete-time state-space equation into STCIM by defining the latent variables with state variables in Eq.~\eqref{eq:state space  equations}. 
Assuming that the latent space $z$ of STCIM is the state variables at time $t-1$ and $t$ ($z_1=u(t)$, $z_2=u(t-1)$ and $z=[z_1,z_2]$) and the input and output of the STCIM are $x$ and $\hat{y}$, respectively. Then, the discrete state-space loss function can be given by
\begin{equation} \label{eq:state space loss function}
\begin{split}
Loss_{state}=\gamma_{1}(\hat{y}-H(z_1,\theta_{3})-J(x,\theta_{4}))+\\\gamma_{2}(z_1-\Phi(z_2,\theta_{1})-G(x,\theta_{2}))
\end{split}
\end{equation}
where $\gamma$ is the weight coefficient. The discrete state-space loss function is used to constrain the latent variables to follow the state-space equations, i.e. the underlying laws of physics. By adding it to the MSE loss, we get the physics-informed interpretable latent space loss function, given by
\begin{equation} \label{eq:physics-informed interpretable latent space loss function}
\begin{split}
Loss=L(y,\hat{y})+ \lambda R(\theta)+ Loss_{state}
\end{split}
\end{equation}

When $\gamma=[\gamma_{1},\gamma_{2}] \to \infty $, the STCIM completely trusts the state-space model. In contrast, when $\gamma=[\gamma_{1},\gamma_{2}] \to 0 $, the STCIM is a data-driven model without physical prior knowledge. {Assuming that the devices follow a unified non-linear model across different operating conditions, the optimal $\gamma$ is certain and determined by whether this model can be trusted,i.e., how simplified the state-space model is.}
By balancing the tendencies towards data-driven and physics-driven models, STCIM can achieve the trade-off in the loss function. According to Eq.~\eqref{eq:physics-informed interpretable latent space loss function}, the physics-informed interpretable latent space loss function consists of four sub-losses: 1) the MSE loss $L(y,\hat{y})$, 2) the parametric regularization of the model $R(\theta)$, 3) the output equation loss $\hat{y}-H(z_1,\theta_{3})-J(x,\theta_{4}$, 4) the state equation loss $z_1-\Phi(z_2,\theta_{1})-G(x,\theta_{2})$. Although the state variables are not measurable, the neural network will output the actual state variables due to the latent space being constrained at both sub-losses 3 and 4, which provides physical properties.

\section{Experimental Results} \label{sec:Simul}
To validate the efficiency of our method, we test the proposed STCIM in three stages. First, analytical experiments based on two generated datasets are used to demonstrate the effectiveness of STCIM. Then, a real-life dataset from a 330MW power plant is utilized as a comparative experiment to further validate the robustness and efficiency of STCIM. The scalability of STCIM is tested using a dataset from 1000MW ultra-supercritical once-through boiler-turbine unit.
At last, we emphasize the importance of position encoding, far topological coupling among variables, and hyper-parameters in the loss function.

\subsection{Experiment Setup}

The two generated datasets include a rotor equation-reduced rotary speed dataset and a 3-order synchronous generator dataset, which are both important for the equipment of thermal power plant. The comparative experiment has 330MW and 1000MW coal-fired power plant datasets that reflects a complex industrial process with heat and mass transformation. All the datasets have two operation conditions, and each condition has at least 50,000 points with noise. The model is trained in one operation condition and tested in another to highlight the effectiveness of STCIM. The state variables are not used as input to models. 

For each STCIM block, we have different parameters. In the local spatio-temporal feature extraction layer block, the features are extracted by two 2D-convolution layers with 8 filters, and each kernel is $3\times 3$. The far topological alignment layer has four heads, and each head has three linear transformations, and each linear layer has 32 neurons. The number of interpretable variables is chosen based on the real physical model. For the decoder, each fully connected layer has 32 neurons and GeLU as an activation function. For all experiments, $\gamma_{1}$ and $\gamma_{2}$ in Eq.~\eqref{eq:state space loss function} are set as 1 and 0.5 separately. During the training process, the latent variables and the target output of STCIM will be integrated into the discrete state-space loss function and then collaboratively optimized by the backpropagation algorithm. 

The STCIM is composed of two sub-blocks with computationally intensive expressions: the local spatio-temporal feature extraction block and the far topological alignment block. For the attribute size of input, $X(t)\in R^{H\times W\times C}$, the computation will intensify in the local spatio-temporal feature extraction block by $\mathcal{O}(K^2 \times C \times M \times H \times W)$ factor, where the K is the kernel size, M is the number of output channels of the block. Similarly, the computation complexity of the far topological alignment block is $\mathcal{O}(D\times B \times H \times (W^2 + H \times W + W \times m))$, where D is the total number of attention layers and feedforward layers, B is the batch size, $m$ is the hidden dimension of the MLP. Since the local spatio-temporal feature extraction block only contains the two 2D-convolution layers, the computational complexity of STCIM is mainly affected by far topological alignment block which contains attention and feedforward layers. The attention mechanism’s $\mathcal{O}(H^2 \times W)$ term may be significant for long sequences, while the $\mathcal{O}(W^2)$ term dominates for large model dimensions.

Six more deep learning-based soft sensor modeling methods are explored for comparisons, including PhyLSTM~\cite{9771395}, PIML~\cite{PRXEnergy}, MSACNN~\cite{10465636}, VCACNN~\cite{10458995}, MIAN~\cite{10648819}, and TimeVAE~\cite{shen2025gaussian}. The six methods include the PINN-based interpretable models and the CNN-based, attention-based, and VAE-based spatiotemporal features extraction models. They have demonstrated their out-performance in different soft sensor of large-scale industrial processes.
Table~\ref{Table: The PRM of all models} shows the number of parameters of models. For each experiment, we train the models with a batch size
of 64 or 128 using the ‘NAdam’ optimizer with an initialized learning rate of 0.01. The learning rate decays by a factor of 0.1 every 30 epochs. The max epoch is set to 100. 
The experiments are implemented in Tensorflow using a CPU Intel i7-11800H CPU at 2.3Hz and an NVIDIA T600 GPU.

\begin{table}[!t]
\centering\footnotesize
\caption{The PRM of main models. \#PRM denotes the number of parameters.}
\label{Table: The PRM of all models}
\renewcommand{\arraystretch}{1.3}
\resizebox{1\linewidth}{!}{\begin{tabular}{ l| c c c c }
\toprule
Models & STCIM & PhyLSTMs~\cite{9771395} & PIML~\cite{PRXEnergy} & VCACNN~\cite{10458995}  \\
\hline
\#PRM & 0.023M  & 0.428M & 0.025M  & 0.371M  \\
\hline
Models & MSACNN~\cite{10465636} &  MIAN~\cite{10648819} & TimeVAE~\cite{shen2025gaussian}  & - \\
\hline
\#PRM & 0.082M  & 0.012M & 0.302M  & -   \\
\bottomrule 
\end{tabular}}
\vspace{-0.3cm}
\end{table}

\subsection{Analytical Experiments} 
In this section, we use two widely used equipment models in thermal power plants
to demonstrate the reliability and scalability of STCIM in modeling and condition monitoring. For the evaluation, we use MAE, MSE and MAPE as the indicators. 

\begin{table*}[htbp]
\centering
\footnotesize
\caption{Evaluation metrics of the 3-order synchronous generator} 
\label{EVALUATION METRICS OF the 3 order synchronous generator}
\renewcommand{\arraystretch}{1.3}
\resizebox{0.98\linewidth}{!}{\begin{tabular}{c|c|c|c|c|c|c|c|c}
\toprule
\multicolumn{2}{c|}{Evaluation}  & STCIM & PhyLSTMs~\cite{9771395} & PIML \cite{PRXEnergy} & VCACNN \cite{10458995} & MSACNN \cite{10465636}& MIAN~\cite{10648819} & TimeVAE~\cite{shen2025gaussian}\\
\hline 
\multirow{4}{*}{MAE ($10^{-2}$)} & $i_{q}$ & \textbf{3.38}  &3.45 &3.57 & 3.41 &3.66 &  3.52 & 3.69 \\
\cline{2-9}
 & $i_{d}$ & \textbf{3.63} &4.07 &4.23 & 3.78 & 4.05 & 3.95 & 4.07 \\
\cline{2-9}
 & $w$ & \textbf{9.18} &16.94 &12.51 & 10.27 & 11.03 & 10.54 & 14.12 \\
\cline{2-9}
 & $\hat{E} _{q}$ & 1.46 &1.45 &1.47 & 1.42 & \textbf{1.33} &  1.40 & 1.42 \\ 
\hline  
\multirow{4}{*}{MSE ($10^{-3}$)} & $i_{q}$ & \textbf{1.75 } &1.79 &1.88 & 1.80 &1.99  & 1.83 & 1.94 \\
\cline{2-9}
 & $i_{d}$ & \textbf{2.08} &2.55 &2.73 & 2.62 & 2.72 & 2.65  & 2.71 \\
\cline{2-9}
 & $w$ & \textbf{13.12} &46.30 & 57.13 & 36.78 & 18.05 & 25.45 & 39.78 \\
\cline{2-9}
 & $\hat{E} _{q}$ & \textbf{0.32} & 0.35 & 0.34 & 0.35 & 0.36 & 0.36 & 0.36 \\ 
\hline  
\multirow{4}{*}{MAPE} & $i_{q}$ & \textbf{4.15\%}  &4.24\% &4.38\% & 4.26\% &4.47\% & 4.34\% & 4.51\% \\
\cline{2-9}
 & $i_{d}$ & \textbf{27.03\%} &28.44\% &29.32\% & 28.01\% & 27.21\% & 27.68\% & 27.46\% \\
\cline{2-9}
 & $w$ & \textbf{9.18\%} & 16.93\% & 17.29\% & 16.87\% & 11.02\% & 15.48\% & 13.98\% \\
\cline{2-9}
 & $\hat{E} _{q}$ & 1.52\% & 1.52\% & \textbf{1.48\%} & 1.51\% & 1.49\%  & 1.50\% & 1.49\% \\ 
\bottomrule 
\end{tabular}}
\vspace{-0.3cm}
\end{table*}

\subsubsection{Rotor Equation-Reduced Rotary Speed}

The stability of the rotor is one of the key factors for the stable operation of thermal power plants. During the operation, 
the rotor needs to maintain a certain speed and stability to ensure the normal operation of the machine and avoid the damage to the machine. 
The reduced rotary speed can be given by
\begin{equation}\label{Rotor}
\scalemath{1}{
\begin{split}
& \frac{\mathrm{d} n}{\mathrm{d} t} =\frac{900}{\pi^{2}Jn}(N_{T}-N_{C}), \\
& \hat{n}=\frac{n}{\sqrt[2]{\frac{T_{1}}{288} } }
\end{split}
}
\end{equation}
where $n$ is the rotor speed, $J$ is the rotary inertia, $\hat{n}$ is the output reduced rotary speed, $T_{1}$ is the air temperature and $N_{T}$ and $N_{C}$ are separately the generated power by the turbines and consumed power by the compressor. The generated dataset contains all the inputs, outputs, and states with 80,000 points with noise.

\begin{figure}[!t] 
\centering
\includegraphics[width=0.95\linewidth]{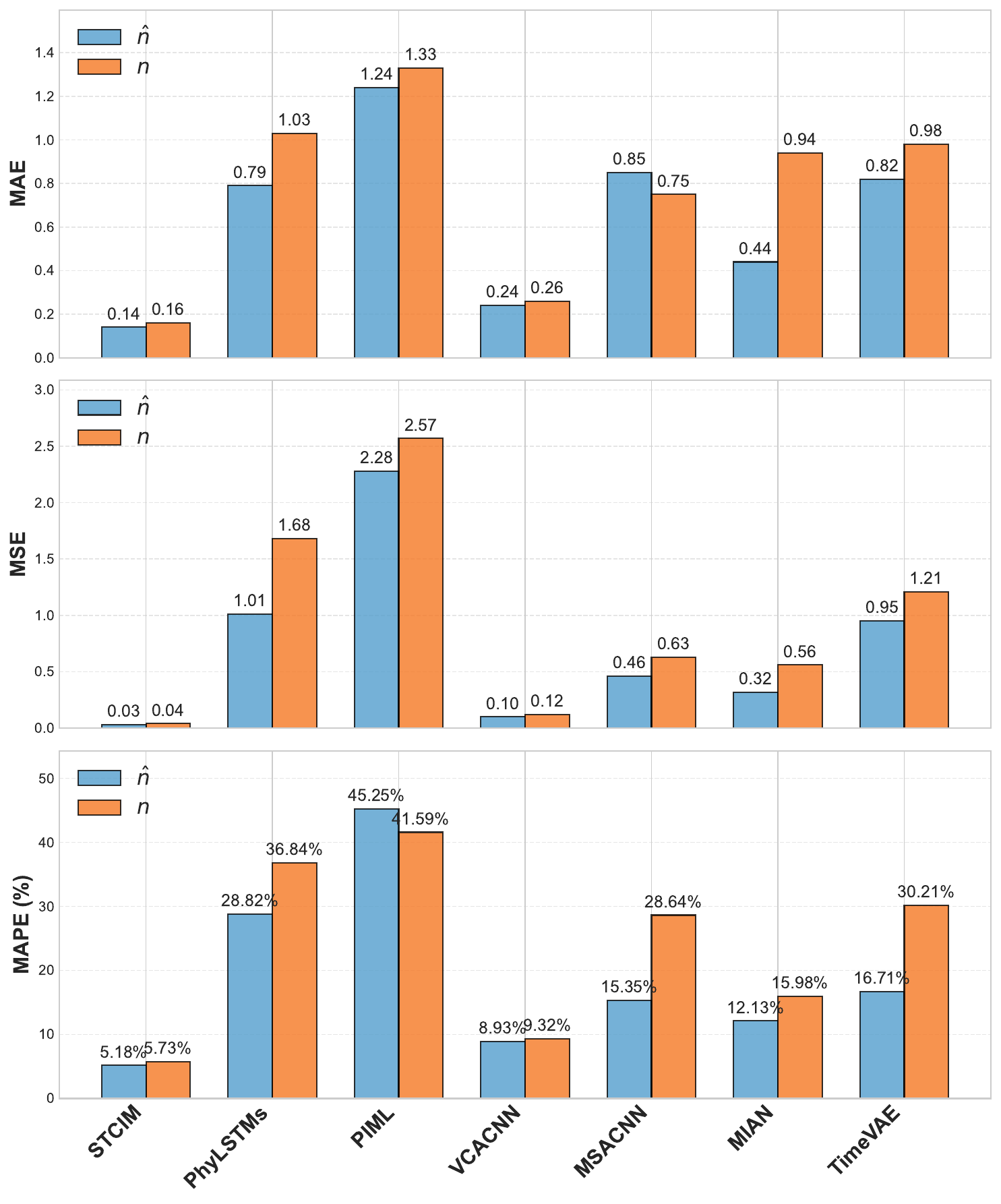}
\caption{Evaluation metrics of rotor equation reduced rotary speed.}
\label{fig:Evaluation_metrics of rotor equation-reduced rotary speed}
\vspace{-0.6cm}
\end{figure}

Fig~\ref{fig:Evaluation_metrics of rotor equation-reduced rotary speed} shows that in a simple equipment monitoring, the STCIM completely outperforms all other models, including the physics-based and feature extraction-based models. First, the conventional fully connected network is not enough to extract the necessary features, resulting in the worst performance of PIML. Then, the time-related features can improve the forecasting precession, which is reflected in PhyLSTM and TimeVAE. Due to the limited variables and the fast characteristics, the local or the global spatio-temporal features in the rotor equation are similar, which makes the performances of VCACNN, MSACNN, and MIAN similar. However, in this experiment, the underlying physical laws in PhyLSTM and PIML do not improve the accuracy of state variable estimation, which may be due to the simplicity of the models.

\subsubsection{3 Order Synchronous Generator}

The synchronous generator is the core equipment in the thermal power system, responsible for converting mechanical energy into electrical energy. Its stable operation is directly related to the reliability and safety of the entire power system. We use the classic 3-order synchronous generator as the simulation model, which is given as follows for data generation.

\begin{equation}\label{Rotor}
\begin{split}
&T_{do}'\frac{\mathrm{d} \hat{E_{q}}}{\mathrm{d} t} =E_{f}-\hat{E_{q}}-(x_{d}-x_{d}')i_{d},\\
&T_{J}\frac{\mathrm{d} w}{\mathrm{d} t}=P_{m}-[\hat{E_{q}}i_{q}-(x_{d}'-x_{q})i_{d}i_{q}],\\
&\frac{\mathrm{d} \delta }{\mathrm{d} t} = w-1,\\
&u_{d}=x_{q}i_{q}-r{a}i_{d},\\
&u_{q}=\hat{E_{q}}-x_{d}'i_{d}-r_{a}i_{q}.
\end{split}
\end{equation}
where $i$ is the input current, $u$ is the output voltage, $w$ is the angular velocity, and $\hat{E_{q}}$ is the transient internal voltage. Since $\delta$ is not related to outputs, it will be ignored in our case. Other parameters are constants. The generated dataset contains all the generated currents, voltages, angular velocity and power with 50,000 points for each operation condition.

Table~\ref{EVALUATION METRICS OF the 3 order synchronous generator} shows the evaluation of the state estimation ($w, \hat{E_{q}}$) and the output regression ($i_{q}, i_{d}$). The proposed STCIM outperforms the other deep learning methods in most evaluation metrics, especially in the output regression and $w$ estimation. Compared to other models, the performance of $\hat{E_{q}}$ estimation of STCIM is competitive as well. The bias in the latent variable $\hat{E_{q}}$ may be caused by the hyper-parameter $\gamma_2$ of loss function.  Considering the good performance of $w$ estimation and that the dataset of $3^{rd}$ order synchronous generator is generated, the hyper-parameter $\gamma_2$ can be set larger for better performance.
Similar to fig~\ref{fig:Evaluation_metrics of rotor equation-reduced rotary speed}, due to the local or global spatio-temporal feature extraction, VCACNN, MSACNN, and MIAN have better performance than PhyLSTM, PIML, and TimeVAE. As the complexity of the model increases, accurate feature extraction becomes increasingly important. Another interesting thing is that the regression accuracy of state estimation ($w, \hat{E_{q}}$) of PhyLSTM and PIML is becoming competitive, which is different from Fig~\ref{fig:Evaluation_metrics of rotor equation-reduced rotary speed}
. We need to emphasize that the regression of state estimation ($w, \hat{E_{q}}$) of PhyLSTM and PIML is unsupervised, since we add the discrete state-space loss in the final loss function like STCIM. They are fundamentally different from VCACNN, MSACNN, MIAN, and TimeVAE.

\begin{table*}[htbp]
\centering
\footnotesize
\caption{Evaluation metrics of the real-life dataset} 
\label{Evaluation metrics of the real-life dataset}
\renewcommand{\arraystretch}{1.3}
\resizebox{0.95\linewidth}{!}{\begin{tabular}{c|c|c|c|c|c|c|c|c}
\toprule
\multicolumn{2}{c|}{Evaluation}  & STCIM & PhyLSTMs~\cite{9771395} & PIML \cite{PRXEnergy} & VCACNN \cite{10458995} & MSACNN \cite{10465636}& MIAN~\cite{10648819} & TimeVAE~\cite{shen2025gaussian}\\
\hline 
\multirow{3}{*}{MAE (Test 1)} & $y$ & 0.45 &0.41 & \textbf{0.40} & 0.46 & 0.78  & 0.51 & 0.58 \\
\cline{2-9}
 & $z_{1}$  & \textbf{0.43} & 0.76 & 0.61  & 0.44 & 1.07 & 0.67 & 0.92  \\
\cline{2-9}
 & $z_{2}$  &\textbf{ 0.18} & 0.21 & 0.69 & \textbf{0.18} & 0.51 & 0.35 & 0.49 \\
\cline{2-9} 
\hline  
\multirow{3}{*}{MSE (Test 1)} & $y$ & 0.31 & 0.29 & \textbf{0.26} & 0.31 & 1.01 & 0.52 & 0.65 \\
\cline{2-9}
 & $z_{1}$  & \textbf{0.25} & 0.87 & 0.59 & 0.29 &  1.73 & 0.46 &  0.98\\
\cline{2-9}
 & $z_{2}$  & \textbf{0.05} & 0.08 & 1.71  & 0.06 & 0.43 & 0.26 & 0.37 \\
\cline{2-9}
\hline  
\multirow{3}{*}{MAPE (Test 1)} & $y$ & 0.37\% & 0.26\% & \textbf{0.25\%} & 0.45\% & 0.78\% & 0.52\% & 0.76\% \\
\cline{2-9}
 & $z_{1}$  & 1.20\% & 1.98\% & 1.58\% & \textbf{1.12\%}  &  2.77\% & 1.69\% & 2.35\% \\
\cline{2-9}
 & $z_{2}$ & \textbf{0.98\% }& 1.12\% & 1.61\% & 0.99\% & 2.66\%  & 1.32\% & 2.03\% \\
\hline 
\multirow{3}{*}{MAE (Test 2)} & $y$ & \textbf{0.46} & 0.54 & 0.57 & \textbf{0.46} & 0.75 & 0.59 & 0.66 \\
\cline{2-9}
 & $z_{1}$  & \textbf{0.11} & 0.56 & 0.14 & 0.28 & 0.68 & 0.32 & 0.58 \\
\cline{2-9}
 & $z_{2}$  &\textbf{0.05} & \textbf{0.05} & 1.38 & 0.11 & 0.26 & 0.18 & 0.22 \\
\cline{2-9} 
\hline  
\multirow{3}{*}{MSE (Test 2)} & $y$ & \textbf{0.49} & 0.50 &0.67 & 1.08 & 0.94 & 0.92 & 0.87 \\
\cline{2-9}
 & $z_{1}$  & \textbf{0.02} &0.07 & 0.03 & 0.05 & 0.07 & 0.07 & 0.06 \\
\cline{2-9}
 & $z_{2}$  & \textbf{0.03} & 0.04 & 0.07 & 0.16 & 0.09  & 0.12 & 0.10 \\
\cline{2-9}
\hline  
\multirow{3}{*}{MAPE (Test 2)} & $y$ & \textbf{0.14\%} & 0.15\% & 0.21\% & 0.48\% & 0.23\% & 0.37\% & 0.29\% \\
\cline{2-9}
 & $z_{1}$  & \textbf{0.35\%} & 0.65\% & \textbf{0.35\%} & 0.44\%  & 0.55\% & 0.48\% & 0.52\% \\
\cline{2-9}
 & $z_{2}$ & \textbf{0.33\%} & 0.34\% & 0.54\% & 0.67\% & 0.61\% & 0.65\% & 0.62\% \\
\bottomrule 
\end{tabular}}
\vspace{-0.3cm}
\end{table*}

\subsection{Comparative Experiments } 

This section is used to demonstrate the effectiveness of STCIM in a real thermal power plant. We collect the dataset from a real 330MW thermal power plant, which follows the introduction in Section~\ref{sec:background}. 
In the dataset, there are two operating conditions and 13 features, and each feature has 96,000 points with noise. The generalization of STCIM is tested under both the same operation condition (Test 1) and a different operation condition (Test 2).

\subsubsection{The comparisons of different methods}
Table~\ref{Evaluation metrics of the real-life dataset} summarizes the evaluation metrics of all models in both Test 1 and Test 2. At the same operating point, the performance of the models varies, which might be caused by the size of models, the different captured features. For models that only focus on local features (MSACNN), the performance is almost the worst when predicting the output and the state variables at the same time. For long time-series data, the 
models related time-related features (PhyLSTM and TimeVAE) outperform MSACNN, indicating the importance of long-term characteristics in the data. In PIML, time as input and domain decomposition help it achieve good performance, especially the output forecasting. Models with spatio-temporal feature achieve the best performance among all models, including STCIM, VCACNN, and MIAN. Benefiting from the built-in physics-informed interpretable latent space loss function, STCIM has more accurate state variable monitoring without labels. 
However, in another operating point, the generalization of STCIM is much better than that of other models. An interesting thing is that the generalization of the physic-informed models (PhyLSTM and PIML) in this experiment is better than other feature extraction-based models, especially the state variables forecasting. It indicates that when the thermal power plant follows unified physical laws, embedding the physical information is equivalent to extracting the intrinsic spatio-temporal features, without complex network architecture, which helps the model to generalize more easily.

\subsubsection{The visualization of output and interpretable latent variables}
\begin{figure}[!t]
\centering
\subfloat[Test 1 of the real thermal power plant dataset]{
\includegraphics[width=0.45\linewidth]{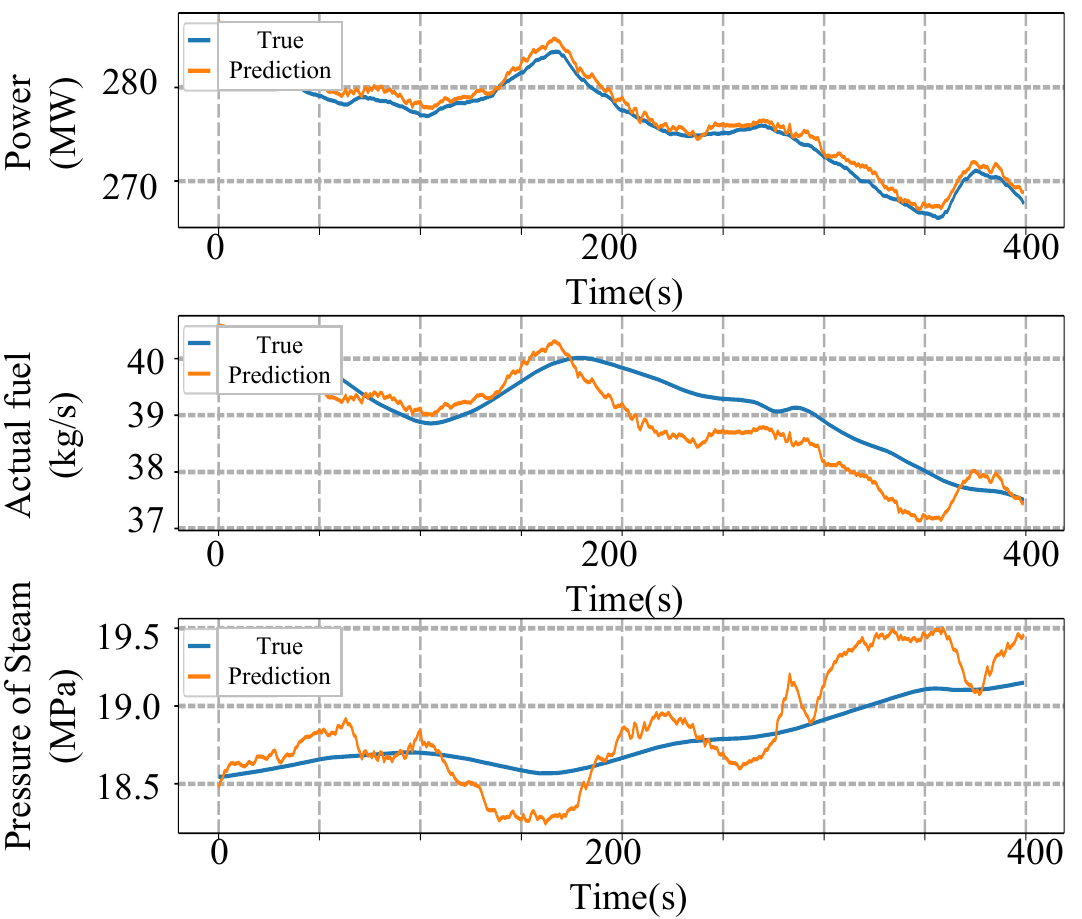}
\label{fig:330MW real data for self}
}~
\subfloat[Test 2 of the real thermal power plant dataset]{
\includegraphics[width=0.45\linewidth]{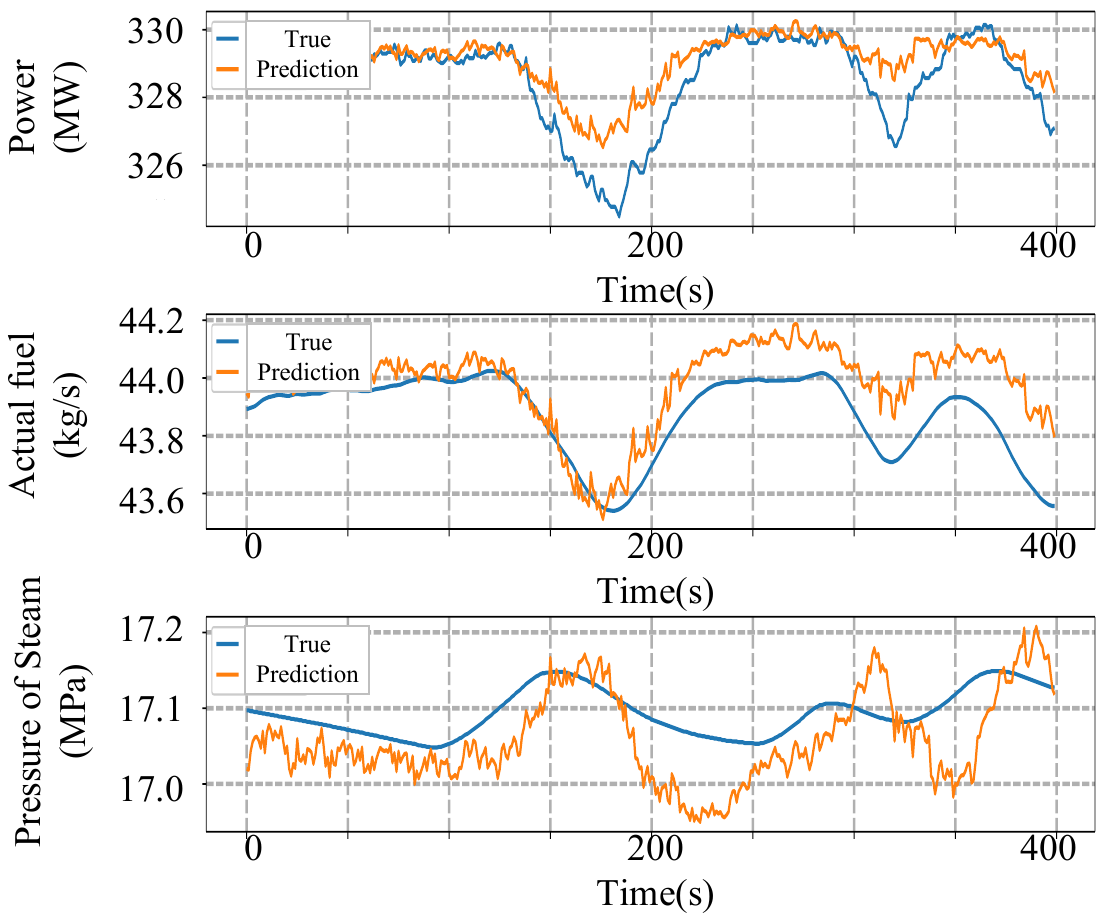}
\label{fig:330MW real data for new}
}
\caption{The output forecasting and state variables monitoring of comparative experiments}
\label{fig:comparative experiments}
\vspace{-0.3cm}
\end{figure}

Fig.~\ref{fig:comparative experiments} shows the output forecasting and state estimation (interpretable latent variables) of the real thermal power plant dataset using STCIM. To reveal the performance of STCIM, 400 points in the test dataset are selected in the figures, including Test 1 and Test 2. Although the performance of STCIM in output forecasting in Test 1 is not the best, it still closely follows the actual value. Compared with Fig.~\ref{fig:330MW real data for self}, the performance of STCIM in Fig.~\ref{fig:330MW real data for new} is better, owing to the obvious fluctuations in state variables, the detailed information from sub-devices,and the physic-informed loss function. Therefore, as well as the spatio-temporal feature consistency, the explicit interpretable latent space along with the physic-informed loss function helps STCIM maintain a good generalization ability in prediction and monitoring under a new operating point.


\subsubsection{The scalability of STCIM: 1000MW Thermal Power Plant}
To demonstrate that STCIM can be applied to other thermal power plants, we use a real dataset from a 1000MW ultra-supercritical once-through boiler-turbine unit~\cite{fan2021dynamic}. The inputs are separately the fuel signal (kg/s), total water flow (kg/s), turbine valve opening (\%). The latent variables ($z_1, z_2$) are separately the steam-water separator pressure (MPa) and the steam-water separator enthalpy (kJ/kg). The outputs ($y_1, y_2$)are the main steam pressure (MPa) and the output power (MW). Each features has 96,000 points with noise.

\begin{figure}[!t] 
\centering
\includegraphics[width=0.95\linewidth]{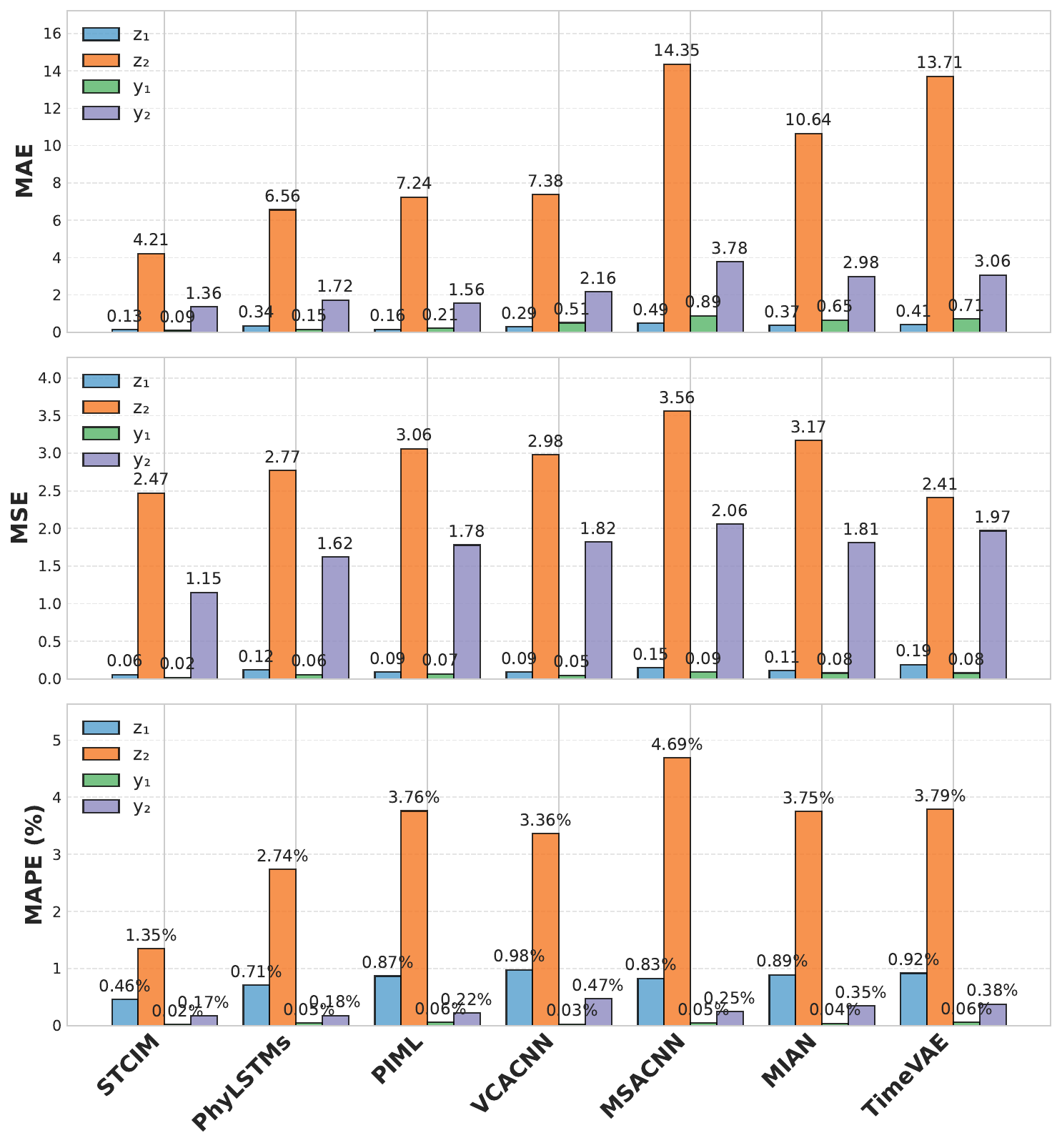}
\caption{Evaluation metrics of 1000MW thermal power plant.}
\label{fig:Evaluation metrics of 1000MW thermal power plant}
\vspace{-0.3cm}
\end{figure}

Fig~\ref{fig:Evaluation metrics of 1000MW thermal power plant} summarizes the evaluation metrics of all models when tested under a different operation condition. STCIM outperforms other models in this experiment. Since the process of the 1000MW ultra-supercritical once-through boiler-turbine unit is more complex than that of the 330MW thermal power plant, the differences between the models are more significant. The integrated physical information loss function forces the models (STCIM, PhyLSTM, and PIML) to satisfy the first principles, which improves the generalization ability of the models, especially in steam-water separator enthalpy forecasting. Instead of extracting spatio-temporal features from a single operation condition, the additional loss and the adopted spatio-temporal consistency can provide more potential physical information to ensure the safety of thermal power plants.

\subsection{Ablation Experiments} 
In this section, we discuss the importance of spatio-temporal consistency and the hyper-parameters of the physics-informed interpretable latent space block in STCIM. The real-life dataset of 330MW thermal power plant is utilized in ablation experiments. The setup is the same as in comparative experiments.

\subsubsection{The importance of adaptive position encoding}


Table~\ref{tab:Evaluation metrics of experiments in real-life dataset without position encoding} shows the evaluation metrics of Test 1 and Test 2 without adaptive position encoding. Compared with Table~\ref{Evaluation metrics of the real-life dataset}, there is a slight decrease in the output forecasting and variable estimation, indicating the effectiveness of spatio-temporal consistency. Whereas, the difference is not remarkable. In contrast, the generalization of STCIM without adaptive position encoding in Test 2 is worse compared to the performance of STCIM in Table~\ref{Evaluation metrics of the real-life dataset}. For example, the MSE in $y$ forecasting without adaptive position encoding almost triples compared with STCIM with adaptive position encoding. The improvement of generalization benefits from adaptive position encoding. By position encoding, the features are endowed with additional information, which improves the expressiveness of STCIM.


\begin{table}[!t]
\centering
\caption{Evaluation metrics of experiments in real-life dataset without position encoding} 
\label{tab:Evaluation metrics of experiments in real-life dataset without position encoding}
\renewcommand{\arraystretch}{1.3}
\resizebox{\linewidth}{!}{\begin{tabular}{c|c|c|c|c|c|c|c|c|c}
\toprule
Datasets & \multicolumn{3}{c|}{MSE} & \multicolumn{3}{|c|}{MAE} & \multicolumn{3}{|c}{MAPE}\\
\hline
Type & $y$ & $z_{1}$ & $z_{2}$ & $y$ & $z_{1}$ & $z_{2}$ & $y$ & $z_{1}$ & $z_{2}$  \\
\hline 
Test 1 & 0.83 & 0.35 & 0.06 & 0.81 &0.47 &0.15 & 0.38\% & 1.20\% &1.12\%   \\
\hline 
Test 2 & 1.37 & 0.06 & 0.06& 0.98 &0.21 &0.09 & 2.90\% & 1.48\% &0.54\%   \\
\bottomrule 
\end{tabular}}
\vspace{-0.3cm}
\end{table}

\subsubsection{The importance of far topological alignment} 

Table~\ref{tab:Evaluation metrics of experiments in real-life dataset without far topological alignment} describes how the evaluation metrics change when the far topological alignment is replaced by a CNN model. 
Different from the comparisons in Table~\ref{tab:Evaluation metrics of experiments in real-life dataset without position encoding}, the performance of STCIM at both Test 1 experiment is greatly affected, especially in the $y$ prediction and $z_1$ estimation, which means that for complex thermal power plants and coupled variables, global and topological features are more important. In Test 2 experiment, the performance becomes worse, due to the change of operating point. The lack of far topological alignment makes the model focus more on local features, while ignoring the potential dependencies of far topological correlations of variables, reducing the accuracy of modeling and monitoring, which is fatal for a large and complex industrial process.

\begin{table}[!t]
\centering
\caption{Evaluation metrics of experiments in a real-life dataset without far topological alignment} 
\label{tab:Evaluation metrics of experiments in real-life dataset without far topological alignment}
\renewcommand{\arraystretch}{1.3}
\resizebox{\linewidth}{!}{\begin{tabular}{c|c|c|c|c|c|c|c|c|c}
\toprule
Datasets & \multicolumn{3}{c|}{MSE} & \multicolumn{3}{|c|}{MAE} & \multicolumn{3}{|c}{MAPE}\\
\hline
Type & $y$ & $z_{1}$ & $z_{2}$ & $y$ & $z_{1}$ & $z_{2}$ & $y$ & $z_{1}$ & $z_{2}$  \\
\hline 
Test 1 & 0.97 & 0.42 & 0.06 & 0.90 &0.51 &0.32 & 0.46\% & 1.38\% &1.45\%   \\
\hline 
Test 2 & 1.78 & 0.13 & 0.09& 1.08 &0.32 &0.09 & 2.93\% & 1.56\% &0.82\%   \\
\bottomrule 
\end{tabular}}
\vspace{-0.3cm}
\end{table}

\subsubsection{Sensitivity analysis of hyper-parameters $\gamma$ in Test 2} 
\begin{figure}[!t] 
\centering
\includegraphics[width=1\linewidth]{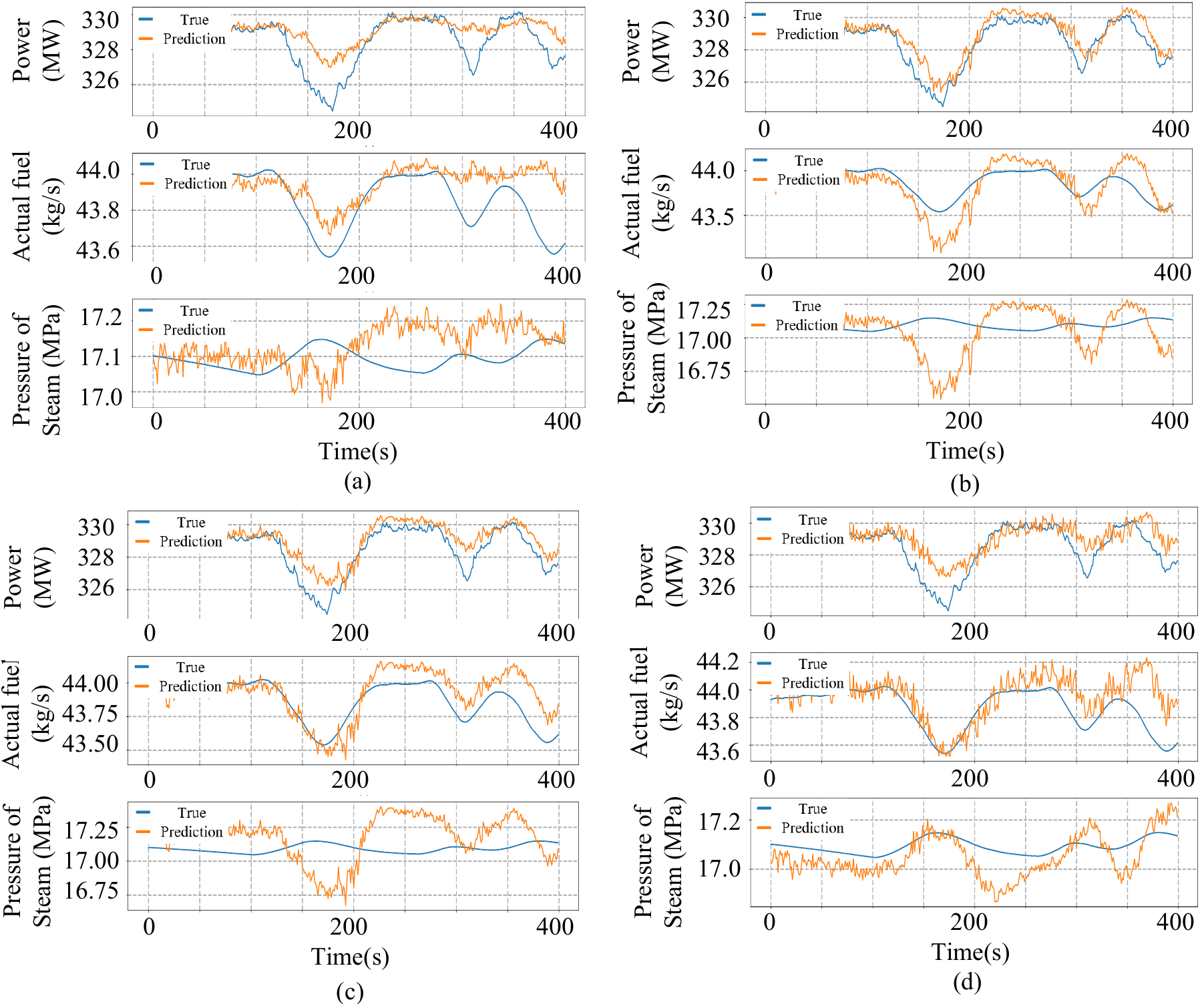}
\caption{Sensitivity analysis of hyper-parameters $\gamma_{1}$ and $\gamma_{2}$ in Test 2 using thermal power plant dataset. (a) $\gamma_{1}=0$ and $\gamma_{2}=0$. (b) $\gamma_{1}=1$ and $\gamma_{2}=0$. (c) $\gamma_{1}=0$ and $\gamma_{2}=1$. (d) $\gamma_{1}=1$ and $\gamma_{2}=1$.} 
\label{fig:Sensitivity analysis of hyperparameters}
\vspace{-0.3cm}
\end{figure}

The impact of hyper-parameters $\gamma_{1}$ and $\gamma_{2}$ on performance of STCIM are tested. Fig.~\ref{fig:Sensitivity analysis of hyperparameters} shows the performance of STCIM using different $\gamma$. When the discrete state-space loss (Fig.(6a)) is not considered in final loss function, STCIM becomes a supervised output and state variables forecasting like VCACNN and the forecasting is unable to align with actual data even though the trend is similar. In Figs.(6b) and (6c), the utilized part of the discrete state-space loss can partly guarantee the fit of state variables. Due to the strong correlation between the state variables and the output, actual fuel and pressure of steam follow the trend of output power, which is different from the actual values. In Fig.(6d), the pressure of steam follows the trend of the real values, due to all the discrete state-space equations are used as part of the loss function.

\begin{table}[!htbp]
\centering
\caption{{Evaluation metrics of experiments with different input dimension}} 
\label{tab:Evaluation metrics of experiments with different input dimension}
\renewcommand{\arraystretch}{1.3}
\scalebox{0.8}{\begin{tabular}{c|c|c|c|c|c|c|c|c|c}
\toprule
{Datasets} & \multicolumn{3}{c|}{{MSE}} & \multicolumn{3}{|c|}{{MAE}} & \multicolumn{3}{|c}{{MAPE}}\\
\hline
{Type} & ${y}$ & ${z_{1}}$ & ${z_{2}}$ & ${y}$ & ${z_{1}}$ & ${z_{2}}$ & ${y}$ & ${z_{1}}$ & ${z_{2}}$  \\
\hline 
{H=5} & {0.82} & {0.06} & {0.11} & {0.62} & {0.22} & {0.13} & {0.28\%} & {0.47\%} & {0.58\%}   \\
\hline 
{H=10} & {0.56} & {0.03} & {0.09} & {0.51} & {0.14} &{0.07} & {0.17\%} & {0.41\%} & {0.40\%}   \\
\hline 
{H=20} & {0.49} &{0.02} &{0.03} & {0.46} &{0.11} & {0.05} & {0.14\%} &{0.35\%} & {0.33\%}   \\
\bottomrule 
\end{tabular}}
\vspace{-0.3cm}
\end{table}

\subsubsection{The impact of input dimension $H$ in Test 2}
{Table~\ref{tab:Evaluation metrics of experiments with different input dimension} discusses the impact of input dimension $H$ on the performance of STCIM ($Input\in R^{H\times W\times C}$). Due to the limited computational resources, we test three different $H$. When the backtracking is small, the advantages of STCIM is hard to see. The inner far topological consistent spatio-temporal feature is replaced by the local spatio-temporal feature, making the performance of STCIM comparable to that of CNN-based models. As $H$ increases, the performance of STCIM is further enhanced. However, considering the computational complexity of STCIM in attention mechanism, a larger $H$ will bring more burden to the device.}

\section{Conclusion}\label{sec:conclusion}

As modern industrial processes become complex, general soft sensor methods are gradually unable to meet the requirements. This paper introduces a novel soft sensor method termed spatio-temporal consistent and interpretable model
(STCIM). We designed a spatio-temporal consistent feature extraction block to capture and align local and global temporal and spatial features from raw data. Subsequently, by giving physical meanings to latent variables and integrating them in a physics-informed interpretable latent space loss function, the latent space maps the spatio-temporal consistent features to an interpretable latent variables to improve the model interpretability. The primary outcome underscores the capability to effectively extract and align the spatio-temporal features and the interpretability of STCIM. The analytical and comparative experiments underscore that STCIM has robust generalization performance across diverse operational conditions, particularly when the training dataset exhibits significant fluctuations in both outputs and state variables, thereby enabling informed and reliable decision-making among power domain experts. The ablation experiments highlight the critical significance of adaptive position encoding, the far topological alignment, and the hyper-parameters of the loss function.

The key limitation of the approach lies in its dependence on prior knowledge of the loss function. When more intricate state-space equations are incorporated into the model, part of the equations can lead to gradient disappearance. Nonetheless, we believe that STCIM has considerable potential for explicitly elucidating the neural network operations. Therefore, there may be significant room for future research, which may focus on the combination of physical models and neural networks.

\end{document}